# A Universal Transition to Turbulence in Channel Flow

Masaki Sano[*] and Keiichi Tamai

[*]e-mail: sano@phys.s.u-tokyo.ac.jp

*Department of Physics, The University of Tokyo*

*7-3-1 Hongo, Bunkyo-ku, Tokyo 113-0033*

**Transition from laminar to turbulent flow drastically changes the mixing, transport, and drag properties of fluids, yet when and how turbulence emerges is elusive even for simple flow within pipes and rectangular channels[1,2]. Unlike the onset of temporal disorder, which is identified as the universal route to chaos in confined flows[3,4], characterization of the onset of spatio-temporal disorder has been an outstanding challenge because turbulent domains irregularly decay or spread as they propagate downstream. Here, through extensive experimental investigation of channel flow, we identify a distinctive transition with critical behavior. Turbulent domains continuously injected from an inlet ultimately decayed, or in contrast, spread depending on flow rates. Near a transition point, critical behavior was observed. We investigate both spatial and temporal dynamics of turbulent clusters, measuring four critical exponents, a universal scaling function and a scaling relation, all in agreement with the (2+1)-dimensional directed percolation universality class.**

Transition to turbulence in open shear flows such as pipe flow and channel flow has been a difficult puzzle for over 130 years[1]. In such flows, the laminar flow becomes turbulent despite its linear stability[5-7]. Also, turbulent structures tend to be localized; laminar states do not break up into turbulent states unless they are invaded by turbulent neighbors. If the tendency for invasion by a turbulent state increases, the turbulent state will eventually spread over the entire space. It is this behavior that led Pomeau to conjecture that the spatio-temporal intermittency observed at the transition from laminar flow to turbulence belongs to the directed percolation (DP) universality class[8-11]. DP is a stochastic spreading process of an active (turbulent) state with a single absorbing state[12], which diverse phenomena such as spreading of epidemics, fires, synchronization[13], and granular flows[14] potentially belong to. Thus, if the transition is continuous and the interaction is short-ranged, then universal critical exponents are expected[12,15]. The linear stability of the laminar flow and recent experimental findings of two competing processes (namely decaying and splitting of a turbulent puff) in pipe flow[16] qualitatively support this analogy[17-20]. However, direct characterization of the transition has been lacking. This situation is presumably due to extremely long time scale of pipe flow, thereby requiring experiments with extraordinarily long pipe to observe the critical phenomena. To overcome this difficulty, we chose a quasi-two dimensional channel flow and forced the inlet boundary condition to be an active (turbulent) state. This enabled us to study the transition to turbulence as a surface critical phenomena. As a result, a clear transition between decay and penetration of the injected turbulent flow was observed. Quantification of the order parameter and the correlation length revealed critical behavior of the transition in the experiment on shear flows; three independent critical exponents support the notion that the transition to turbulence in channel flow belongs to the DP universality class.

In channel flow, the Reynolds number ($Re$) is defined as $Re = Uh/\nu_K$ where $U$ is the centerline velocity of the parabolic profile, $h$ is the half-height of the channel, and $\nu_K$ is the kinematic viscosity of the fluid (Fig. 1). Laminar channel flow (plane Poiseuille flow) is linearly stable up to a Reynolds number of $Re_L = 5772$ [21]. However a turbulent spot excited by a finite perturbation grows and splits to spread into extended spatial regions[22] because of a global nonlinear instability even if $Re$ is much smaller than $Re_L$. To study this transition, an experimental setup is configured. The flow channel has a length of 5880 mm in the streamwise ($x$) direction, a cross section of 5 mm in depth (the $y$ direction), a width of 900 mm in the spanwise ($z$) direction. Thus the aspect ratio of the channel was $2352h \times 360h \times 2h$. The flow dynamics in the ($x,z$) plane were visualized and recorded using a visualization technique and three CCD cameras (see Supplementary Information (SI)). Instead of triggering turbulent spots by a local perturbation for each measurement as in the previous experiments[16,22], turbulent flow is continuously excited in the buffering box via the use of a grid and injected from the inlet ($x = 0$), otherwise the flow remained laminar up to much higher Reynolds numbers (see SI). Figure 1 shows the visualization of turbulent spots observed near the middle of the channel ($x/h = 1200$) at $Re$ =810. Note that most part of the turbulent flow injected at the inlet quickly decayed and changed into a laminar flow. Hence, surviving turbulent flow tends to be depicted by localized turbulent spots characterized by finer scale disordered eddies surrounded by several streaks and clear laminar flows[22]. The typical size of the turbulent spot is about $40 \sim 80h$.

Figure 2 shows normalized intensity images of the flow pattern for three different Reynolds numbers. As shown in Fig. 2a for $Re$=798, the injected turbulent structure separated into localized turbulent spots which gradually decayed as they propagated with the mean flow, and ultimately disappeared before reaching the channel exit. For $Re \geq 830$, splitting and spreading of turbulent spots were clearly observed (see Fig. 2b for *Re*=842). These processes contributed to the creation of turbulent clusters whose dynamics exhibited an intermittent stochastic nature in space and time. For sufficiently large *Re* values (e.g. $Re > 900$), turbulent flow was sustained (see Fig. 2c for *Re*=1005).

This setup enabled the attainment of a steady state measurement of the area fraction of the turbulent region (the turbulent fraction ρ) for various *x*. ρ, estimated by measuring the time fraction occupied by turbulent flow averaged over a protractive time period (approximately 40 min; *i.e.* 100 times of that of flow circulation time), was found to saturate for higher *Re* and for larger *x*, as shown in Fig. 3a. Therefore, the turbulent fraction was measured as a function of *Re* at several distant locations, *x*, satisfying $x/h > 1280$ (see Fig. 3a). The area fraction of the active (turbulent) region is the order parameter in the DP transition which increases continuously from zero to positive values. Thus, the curves are fit by the function, $\rho = \rho_0 \varepsilon^\beta$ in the inset of Fig. 3a, where ε is the reduced Reynolds number, $\varepsilon \equiv (Re - Re^c)/Re^c$. As a result, β =0.58(3) and $Re^c = 830(4)$ were obtained as the best fit values. The value of β was very close to the universal exponent of (2+1)-dimensional (*i.e.*, two-dimensional in space and one-dimensional in time) DP, $\beta^{DP} = 0.583(3)$. Although non-vanishing order parameter below $Re^c$ exists as usual for finite systems, clear critical behavior was observed despite its small effective size. Note that relatively small systems can show remarkably clear power-law behavior in numerical models exhibiting a DP transition (see SI for simulation). Furthermore, the result $Re^c = 830(4)$ is consistent with the results of direct numerical simulations (DNSs) for a channel flow, in which the global instability was reported as $Re^c < 840$ [23].

Moreover, spatial variations of ρ(*x*) over space were investigated. The turbulent fraction ρ(*x*) showed clear exponential decays for *Re* values smaller than 803, while ρ showed saturations at constant values in space for *Re* = 904 as shown in Fig. 3b. Hence, transition between decay and penetration is evident. Thus, we fit ρ(*x*) with an exponential decay; $\rho(x) \sim \exp(-x/L)$ for the data taken at *Re*<*Re*$^c$. The decay length *L* increased as $Re^c$ was approached with a power law, $L \sim |\varepsilon|^{-\nu}$. As a result, ν = 1.1(3) was obtained as the best fit (see the inset of Fig. 3b). This value is close to the critical exponent characterizing the divergence of temporal correlation length, $\nu_\parallel^{DP} = 1.295(6)$. For temporal correlation length $\xi_\parallel$ and spatial correlation length $\xi_\perp$, the relations $\xi_\parallel \sim \varepsilon^{-\nu_\parallel^{DP}}$ and $\xi_\perp \sim \varepsilon^{-\nu_\perp^{DP}}$ hold respectively in DP. As ε approaches to 0, the spatial correlation from the active wall becomes irrelevant compared with the temporal correlation due to the relation; $\nu_\parallel^{DP} > \nu_\perp^{DP}$ (and thereby $\xi_\parallel \gg \xi_\perp$, for $\varepsilon \ll 1$), which holds in DP. Thus, the examination of spatial variation of ρ(*x*) is actually equivalent to the examination of a quenching dynamics of turbulence injected from the inlet that is conveyed to downstream by the flow. This is the very reason $\nu_\parallel$ was observed instead of $\nu_\perp$ from $\rho(x)$. Therefore, the decay length $L$ coincides with the survival length of the active cluster which defines temporal correlation length $\xi_\parallel$ in DP[24-26].

There are three independent exponents that characterize the DP universality class: $\beta$, $\nu_\parallel$ and $\nu_\perp$. Numerical simulation on a simple directed bond percolation model with advection indicates that one can estimate the remaining exponent $\nu_\perp$ by measuring distributions of the durations $\tau$ of the laminar state (laminar interval distribution) $N(\tau)$ at fixed downstream locations for $Re > Re^c$ (see Fig. S6 and Fig. S7). Therefore, the distributions $N(\tau)$ at the $x=3200$ mm were accumulated for 40 different $z$-positions within a half span width ($\pm 225$ mm) around the mid-height. For small $\tau$ values, power law distribution is expected near $Re = Re^c$ reflecting the scale invariance of critical clusters[15]. Figure 4a shows the resulting $N(\tau)$ functionality for several different Reynolds numbers. We fit this by the power law $N(\tau) \sim \tau^{-\mu}$, with $\mu = 1.25(5)$ which is close to the universal exponent in DP, $\mu_\perp^{DP} = 1.204(2)$.

To observe the tail of the distributions, a complementary cumulative probability, $P(\tau) \equiv \int_\tau^\infty N(t) dt / \int_0^\infty N(t) dt$ was calculated and provided in Fig. 4b. We defined the correlation length $\xi$, by fitting the tail of $P(\tau)$ with an exponential function, $P(\tau) \sim \exp(-\tau/\xi)$. As the transition point ($Re^c$) was approached, $\xi$ substantially increases (Fig. 4c). Thus a best fit was determined per $\xi \sim \varepsilon^{-\nu}$ with an exponent $\nu = 0.72(6)$ for a small $\varepsilon$ region ($0.005 < \varepsilon < 0.06$) in accordance with $\nu_\perp^{DP} = 0.733(3)$. Although the range of the power law is limited due to finite size of the system, obtained exponents of $\mu_\perp$, $\beta$, and $\nu_\perp$ consistently satisfy the universal scaling relation of $\mu_\perp = 2 - \beta/\nu_\perp$. As such, these results encourage the further exploration of universal features for the subject phenomena. Thus, a universal scaling hypothesis $P(\tau) \sim \varepsilon^{\nu_\perp(\mu_\perp - 1)} g(\varepsilon^{\nu_\perp} \tau)$ for $P$ is introduced (see SI for a detailed discussion and numerical validation of this hypothesis) with a universal scaling function $g(x)$. By plotting rescaled probability $\varepsilon^{-\nu_\perp(\mu_\perp - 1)} P(\tau)$ as a function of rescaled duration $\varepsilon^{\nu_\perp} \tau$, we find that several curves overlap (see Fig. 4d) when we choose $Re^c = 830$ in accordance with the previous result shown in Fig. 3a. All these results support that the transition can be understood as the DP process conveyed to downstream by the flow.

In conclusion, the present result strongly supports the notion that the transitions to turbulence in shear flows belong to the DP universality class (Obtained critical exponents are summarized in Table 1). Unveiling the "dynamical origin"[28-30] of the critical behavior reported here is a future challenge toward a deeper insight into the onset of turbulence.

**Table 1. Summary of critical exponents measured in this experiment.**

| (2+1)D System | $\beta$ | $\nu_\perp$ | $\mu_\perp$ | $\nu_\parallel$ |
|---|---|---|---|---|
| Channel Flow (present exp.) | 0.58(3) | 0.72(5) | 1.25(5) | 1.1(3) |
| DP Theory | 0.583(3) | 0.733(3) | 1.204(2) | 1.295(6) |

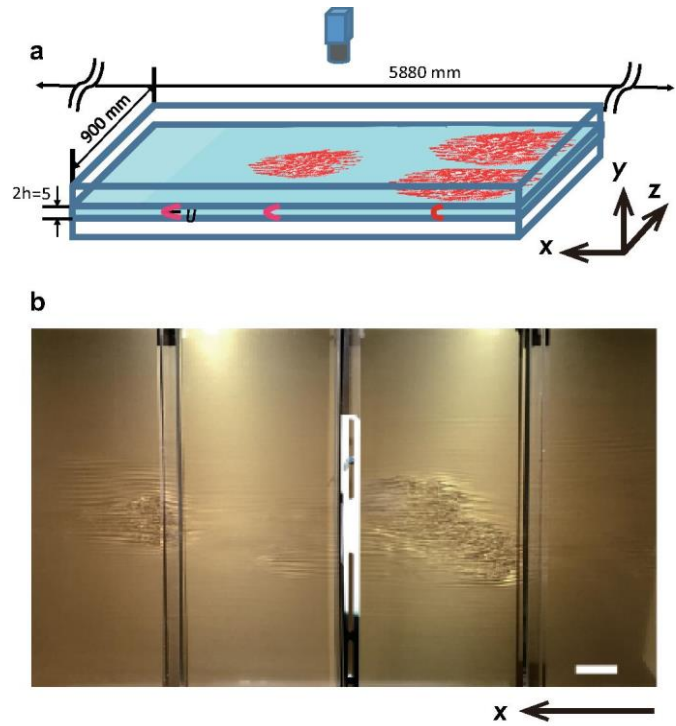

**Figure 1. Apparatus and snapshot of turbulent spots. a**, Schematics of the apparatus. The aspect ratio of the channel is $2352h \times 360h \times 2h$, here the depth $2h$ is 5mm. **b**, Turbulent spots are visualized near the middle (*x*=3 m) downstream location of the channel at *Re*=810. The turbulent flows are injected at the inlet ($x = 0$) of the channel by using a grid. Visualization was made by micro-platelets and grazing angle illuminations. The scale bar is 100 mm.

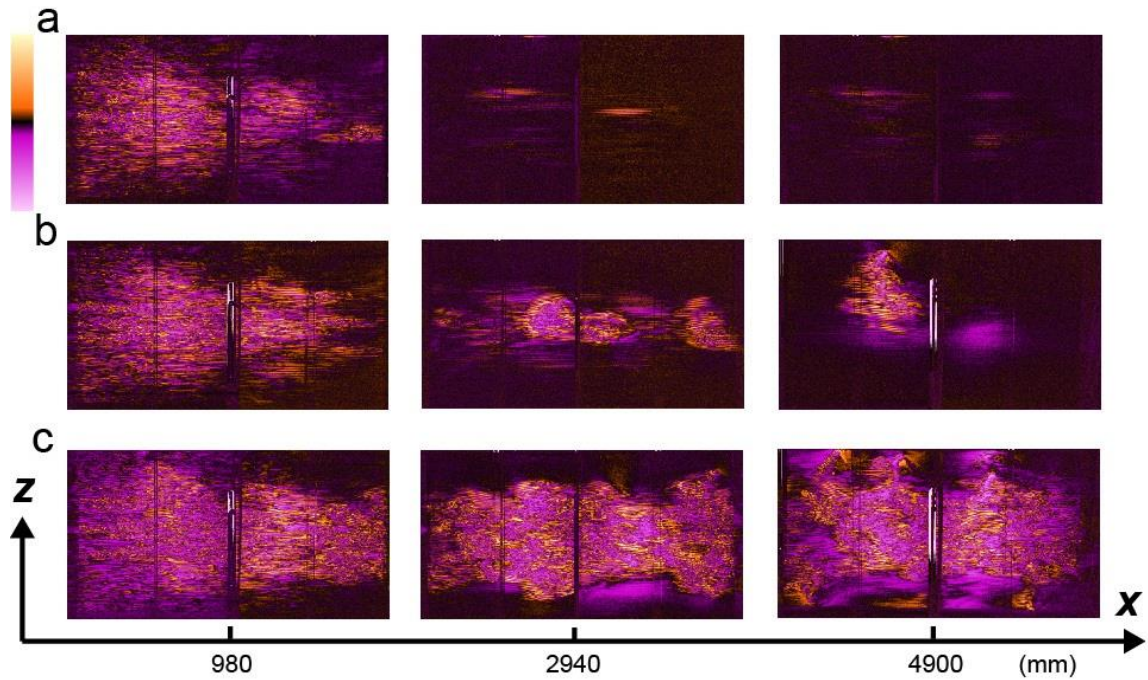

**Figure 2. Spatial variations of the flow across the transition a**, Snapshots taken by three CCD cameras respectively from left to right at *Re*=798. Quick decay of turbulent flow is evident. Color represents normalized image intensity. Black color is assigned to the point where image intensity is close to the laminar state (see the color map.) **b**, Snapshots at *Re*=842. Intermittent nature of the turbulent spots can be seen. **c**, Snapshots at *Re*=1005. Saturation of turbulent fraction is evident.

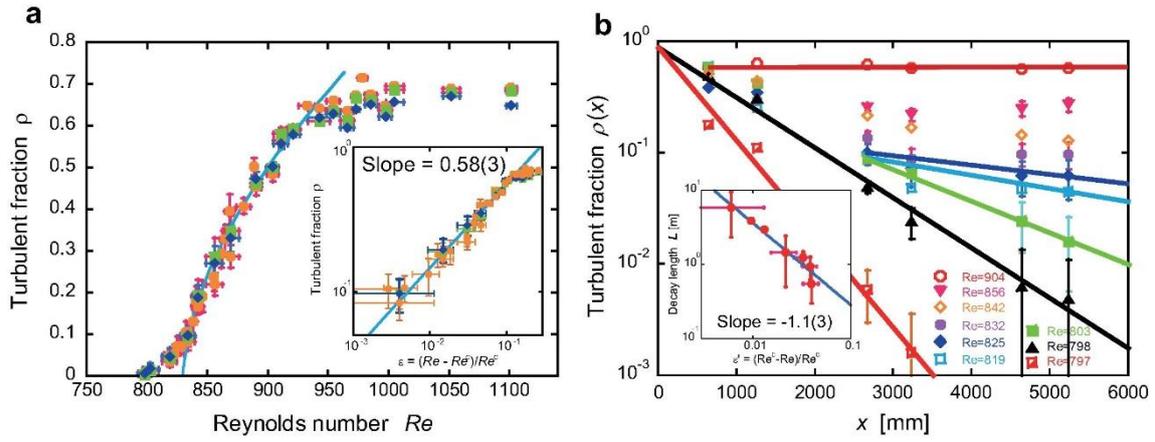

**Figure 3. Critical behavior of the turbulent fraction. a**, The turbulent fraction $\rho$ vs. *Re* is plotted at different downstream locations: $x/h = 1292$ (●), $x/h = 1880$ (◆), and $x/h = 2096$ (■). Inset: A log-log plot of $\rho$ as a function of reduced Reynolds number $\varepsilon$, $\varepsilon \equiv (Re - Re^c)/Re^c$, with $Re^c = 830(4)$. The solid blue lines are the best fit, $\varepsilon^{0.58}$, for the data in $10^{-3} < \varepsilon < 10^{-1}$. **b**, The turbulent fraction as a function of distance *x* from the inlet where turbulence is created by a grid. Measurements were performed for six different *x* locations where the incident angles and the reflection angles of the light are in identical condition. The solid lines are exponential fittings, $\rho(x) \sim \exp(-x/L)$, applied for the data satisfying $x/h > 1040$. Inset: log-log plot for $L$ vs. $\varepsilon' \equiv (Re^c - Re)/Re^c$. Solid line is the best fit, $L \sim |\varepsilon'|^{-\nu}$ with $\nu = 1.1(3)$.

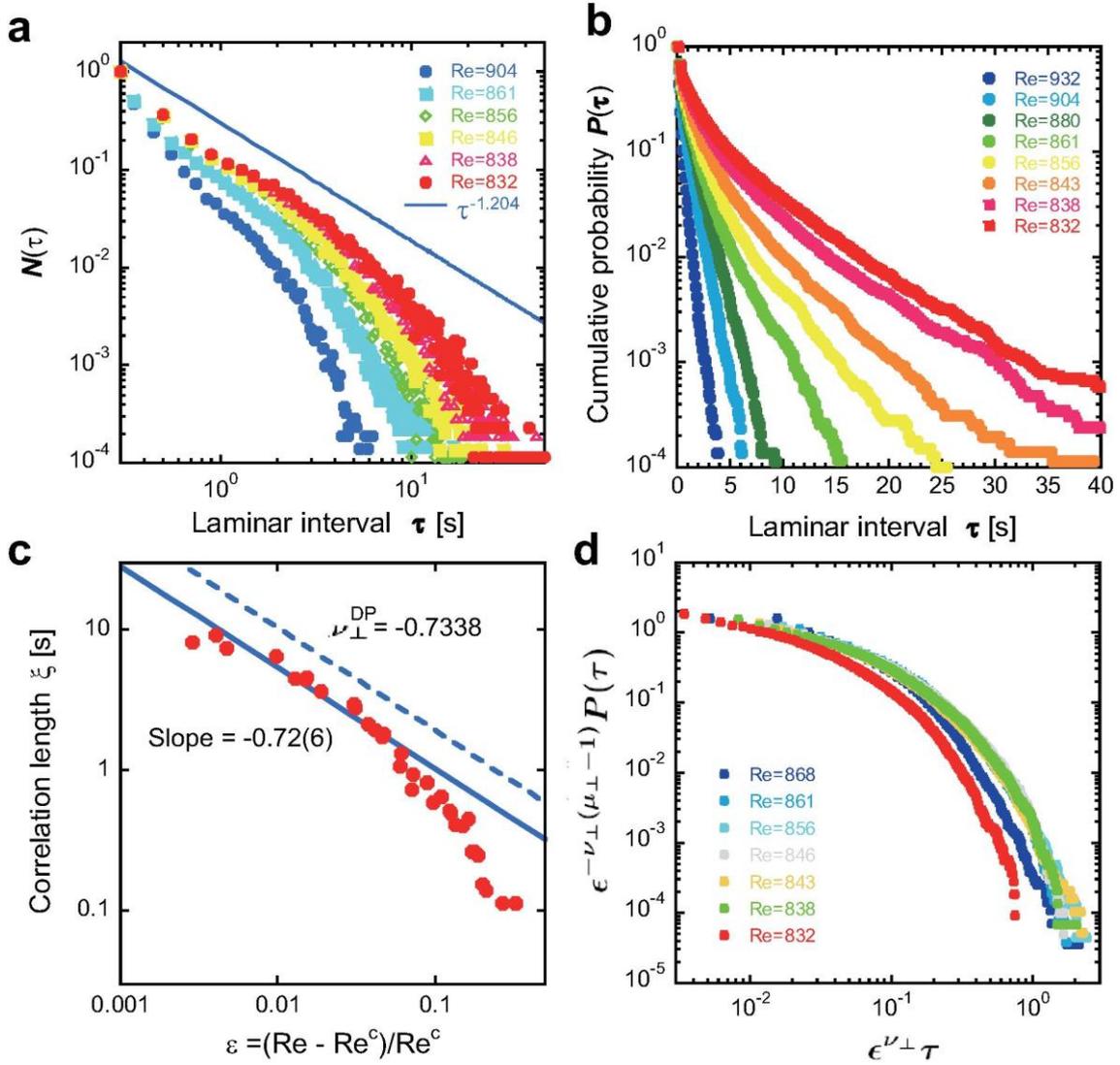

**Figure 4. Critical behavior of correlation length and universal scaling. a**, Laminar interval distribution $N(\tau)$ measured at a fixed downstream location, $x/h = 1280$, for different *Re*. Power law fit for *Re*=846 to 832 approaches to $N(\tau) \sim \tau^{-\mu}$, with $\mu = 1.25(5)$. The solid blue line corresponds to $\mu_{\perp}^{DP} = 1.204$ for eye-guide. **b**, Complementary cumulative laminar interval distribution $P(\tau)$. The tails show exponential decays: $P(\tau) \sim \exp(-\tau/\xi)$. **c**, The correlation length $\xi$ vs. reduced Reynolds number $\varepsilon$. The solid line is the fit, $\xi \sim \varepsilon^{-\nu_{\perp}}$, with $\nu_{\perp} = 0.72(6)$. The broken line corresponds to $\nu_{\perp}^{DP} = 0.733$ for eye-guide. **d**, Data collapse for several different *Re* according to the scaling hypothesis. For $Re^c$, the value estimated in the experiment was used. For $\nu_{\perp}$ and $\mu_{\perp}$, the theoretical values for (2+1)-dimensional DP were used.

**Acknowledgements**

Authors would like to thank M. Kuroda, K. A. Takeuchi, and H. Brand for stimulating discussions. This work is supported by KAKENHI (No. 25103004, "Fluctuation & Structure") from MEXT, Japan, and the JSPS Core-to-Core Program "Non-equilibrium dynamics of soft matter and information".

# Supplementary Information

## Methods

**Construction of the flow channel.**

The channel walls were made of 25-mm-thick Polymethyl methacrylate (PMMA) glass plates of optical surface quality. The entire 6 m (5880mm) channel comprised 3 pieces of 1960 mm×1000 mm slots (see Fig. S1). Both ends of each slot were reinforced by welding 50-mm-thick flanges to ensure the precision of joint between two slots using O-ring. The side walls were made of PMMA strips of 50 mm×1000 mm×5 mm. When constructed in this way, the precision of the depth was ±0.1 mm. To avoid further deflection due to static pressure load in the channel, cross-braces were placed at 425-mm intervals along the channel. The working fluid is water. The channel inlet was connected to a buffering box via a smoothly curved contracting joint whose area contraction ratio was 1 : 20. To set a turbulent boundary condition, we placed a grid near the inlet. (When the grid is covered with 7 layers of mesh screens, the flow remained laminar in a whole channel at least up to $Re$=1400. Since the covering by mesh screen was not sufficient at the edge, turbulent flows did not decay near the both ends of the buffering box near $z$=0 mm and $z$=900 mm at $Re$=1400. Those turbulent flows injected from the inlet gradually grew and spread. Even in that case, there was no spontaneous nucleation of turbulent spot from the laminar state in the middle of the channel at $Re$=1400.) Velocity control was attained by electronically controlling the speed of pump and the opening of the valve. The flow rate was measured by a flow meter (FD-UH40G, Keyence). The temperature of the water was controlled at 25 °C within the accuracy of ±0.1 °C.

**Visualization.**

Since the measurement of the spatio-temporal dynamics of turbulent spots in a large space is problematic, we utilized a simple visualization using tracer particles. Metal coated mica platelets (10 – 20 μm in diameter and 3 μm in thickness, Iriodin, Merck) were added to water for visualization. The concentration of the tracer was reduced to 0.04% in weight to keep the change of viscosity negligible (<0.1% according to Einstein's law[23]). Thin platelets tend to align perpendicular to shear stress which is parallel to the *x-z* surface in laminar flow states while they rotate in turbulent spots. The grazing angle illumination brought moderate light reflections from the laminar regions to the front, while scattering from the turbulent spots is omni-directional and its intensity deviated significantly from that of laminar regions (see Fig. S3). Six projectors (PJ4114NW, 3000 lumen, Ultra short focal length, Ricoh) were attached 300 mm above and 250 mm apart from the (*x,z*) surfaces to illuminate the channel surface with a grazing angle to attain reasonably uniform intensity of illumination. Three CCD cameras (1608 pixels × 1208 pixels, 10 frames/sec) facing the center of the *x-z* plane of each slot synchronously captured movies of spatio-temporal dynamics of the flow of each slot. For the evaluation of the turbulent fraction, ρ(*x*) was measured at 6 positions (*x*=0.65 m. 1.27 m, 2.68 m, 3.23 m, 4.70 m, 5.24 m) where the incident angle from each of 6 light sources to each measuring position in the channel are almost equal, simultaneously the reflection angle from the measuring position to each of 3 CCD cameras are almost equivalent. This choice was made to avoid unwanted inhomogeneity in the turbulent fraction ρ(*x*) due to anisotropic effect of the scattered light from the platelets.

**Image analysis.**

To obtain a background profile, the spatial profile of the scattered light intensity for the laminar state, $R(x,z,t)$, was captured at $Re \sim 600$ as movies for 100 sec every day before and after the series of measurements of turbulence. Then time average of the background $R_0(x,z) = \langle R(x,z,t) \rangle_t$ at each pixel was calculated presuming that the spatial resolution is sufficient. (1 pixel size in the image corresponds to ~1.25 mm (=*h*/2) on the surface of the channel.) After taking movies for each *Re* number, each frame of the movie was normalized by dividing by the background to reduce inhomogeneity due to the illumination, *i.e.*, a normalized image $N(x,z,t)$ was calculated as $N(x,z,t) \equiv I(x,z,t)/R_0(x,z)$, where $I(x,z,t)$ is the original image and $R_0(x,z)$ is the background image. (Figures S2a, S2b and S2c show examples of the original image, $I(x,z,t)$, the background image $R_0(x,z)$, and the normalized image $N(x,z,t)$, respectively as a snapshot.) Then $N(x,z,t)$ is multiplied and shifted in order to save the memory storage. The standard deviation $\sigma(x,z)$ of the intensity fluctuations for the laminar state, $R(x,z,t)/R_0(x,z)$ was calculated. (Figure S3 shows the probability distribution function (PDF) of intensity fluctuations of a laminar state and PDF of the flow with turbulent spots embedded in a laminar state.) Turbulent spots were detected by the following procedure: Time series of the image intensity measured at each pixel point, $N(x,z,t)$, are created by fixing *x* and *z*. If the intensity deviated from expected values of the laminar state more than $\pm n \times \sigma(x,z)$ ($n=3$ is used in most cases), space-time point $(x,z,t)$ is regarded as a turbulent state, otherwise regarded as a laminar state. After assigning turbulent (active) or laminar (inactive) state for all *x*, *z*, and *t*, the cluster sizes of the spatially connected turbulent regions were calculated at each time instance. Clusters whose sizes were larger than $h^2$ were regarded as turbulent regions and the remaining small clusters as laminar regions, by assuming that the minimum size for localized turbulent eddies at moderate *Re* cannot be smaller than the gap size *h*. A typical result of the binarization is shown in Fig. S2d.

**Supplementary Note**

This supplementary note demonstrates the influence of an advection and an active boundary condition on systems exhibiting the transition which falls onto the directed percolation universality class (DP) by a numerical simulation on a simple model.

The model we consider is similar to the (1+1)-dimensional (i.e., one-dimensional in space and one-dimensional in time, hereafter abbreviated as (1+1)D) directed bond percolation [1] model (which is known to exhibit a continuous transition belonging to DP), except the existence of the wall and asymmetry in connection in order to mimic the effect of advection (Fig. S4a). More formally, the model with a percolation probability $p$ is defined on a set $s = \{s_i\}_{i=0}^{N-1}$ of a local binary variable $s_i$ (where $i$ is the number of sites from the wall and $N$ is the size of the system) by the following rules:

$$s_0(t) = 1 \text{ for } \forall t \quad (1)$$

$$s_i(t+1) = \begin{cases} 1; & \text{if } (s_{i-1}(t) = 1 \text{ and } z^- < p) \text{ or } (s_i(t) = 1 \text{ and } z^0 < p) \\ 0; & \text{otherwise} \end{cases} \text{ for } i \geq 1 \quad (2)$$

where $s_i = 0$ and $s_i = 1$ denote an inactive state and an active state respectively, and $z^-, z^0 \in (0,1)$ are randomly generated variables. Note that the critical percolation probability $p_c$ of the model is expected to be identical to the one of ordinary directed bond percolation, which is very accurately estimated to be $p_c = 0.64470015(5)$ in past studies [2]. Typical dynamics of the model is shown in Fig. S4b. We can clearly see (especially for $p = 0.64245$ and $p = 0.64570$) that localized clusters of active sites are moving along the direction of the advection.

We performed a Monte—Carlo simulation on the lattice of the size $N = 8192$. We started our simulation with the system whose site in the wall is active and other sites are inactive. Each Monte—Carlo step consists of a parallel update of each site based on the rule (1) and (2). Here we perform a stationary simulation: We first run the simulation for $15 \times 10^4$ steps (which is about 20 times longer than the steps needed for an active state generated on the wall to reach the other end of the lattice) and then the statistics are accumulated over another $85 \times 10^4$ steps. The simulation is repeated over 16 times to improve statistics.

We first measured the order parameter $\rho$ as a function of the distance $x$ from the wall. The order parameter $\rho(x)$ is defined as a probability that the $x$ th site from the wall is active during the stationary simulation. Figure S5a shows that $\rho(x)$ decays exponentially when the percolation probability $p$ is smaller than the critical value $p_c$, while it saturates to a finite value when $p > p_c$. This behavior allows us to define a decay length $L$ as a characteristic length of the exponential decay at $p < p_c$, i.e., $\rho(x) \sim \exp(-x/L)$ for sufficiently large $x$. As shown in Fig. S5b, we find that $L$ as a function of $p_c - p$ shows a power-law behavior in a vicinity of the critical point. Fitting by a function $L \sim (p_c - p)^{-\nu}$ yields $\nu = 1.71(5)$, which is close to the theoretical value of $\nu_\parallel$ for (1+1)D DP: $\nu_\parallel^{DP} = 1.733847(1)$ (Note that, as usual for a finite system, deviation from the power-law behavior is observed at the point where the resulting $L$ is comparable to $N$; then a finite- size effect is no longer negligible). These results are consistent with those reported by Costa *et al.* [3], who studied asymmetric contact process (which is also known to belong to the DP universality class [4]) driven by an active boundary condition. We also measured the order parameter $\rho$ at the fixed observation point (8,000$^{th}$ site from the active wall) for various value of $p$. As expected, a power-law behavior $\rho(p) \sim (p - p_c)^\beta$ is observed for $p \geq 0.64570$ (Fig. S5c). We obtain the critical exponent $\beta$ as a best fit $\beta = 0.271(8)$, which is in a very good agreement with (1+1)D DP $\beta = 0.276486(8)$, although a deviation from the power law is present for $\varepsilon \sim 10^{-3}$ as expected.

Next, we measured a distribution of durations $\tau$ of an inactive state for a fixed observation point, as done in the experiment discussed in a main text. We made a histogram for $\tau$, and then calculated a complementary cumulative probability distribution $P(\tau)$: $P(\tau) \equiv 1 - \sum_{t=1}^{\tau} N(t) / \sum_{t=1}^{T} N(t)$, where $T$ is a Monte—Carlo steps for each realization. We confirmed that the distribution $P(\tau)$ as a function of $\tau$ converged as long as we measured $P(\tau)$ at the site sufficiently far from the wall for $p > p_c$. The distribution $P(\tau)$ is measured at a fixed point ($x = 5000$) for various values of $p(> p_c)$ and the results are shown in Fig. S6a, where we can clearly see an exponential decay $P(\tau) \sim \exp(-\tau/\xi)$ for a sufficiently large $\tau$. The characteristic length $\xi$ of the decay turned out to obey a power law $\xi \sim (p - p_c)^{-\nu}$, whose exponent is significantly different from the exponent $\nu_\parallel$ associated with a correlation time but very close to the exponent $\nu_\perp$ associated with a correlation length (Fig. S6b). As discussed in the main text, this behavior can be understood by comparing the growth of the correlation length and correlation time. In systems exhibiting the DP universality, the correlation time grows much faster than the correlation length (recall $\nu_\parallel > \nu_\perp$), and therefore the clusters of the active state become thinner and thinner as we approach the critical point (as we can see in Fig. S4b). Thus, by fixing the observation point and measuring the duration of the inactive state, we are actually probing the spatial distance between the clusters, which is expected to correspond to the spatial correlation length.

We now explore the universal scaling hypothesis of the inactive interval distribution. The discussion in the previous paragraph implies the inactive interval distribution probes spatial interval between the active clusters, so we postulate that the inactive interval distribution has scaling properties analogous to that of spatial two-point correlation function. That is, we assume that under the transformation of the control parameter $\varepsilon \mapsto \lambda \varepsilon$ and interval $\tau \mapsto \lambda^{-\nu_\perp} \tau$, the inactive interval distribution also scales as $N \mapsto \lambda^\eta N$ with some suitable exponent $\eta$:

$$N(\tau) \sim \lambda^{-\eta} f(\lambda^{-\nu_\perp}\tau; \lambda\varepsilon), \qquad (3)$$

although the inactive interval distribution is significantly different from an ordinary two-point correlation function as pointed out by Hinrichsen [5]. Accepting the postulation and choosing the scaling parameter $\lambda = \tau^{1/\nu_\perp}$, simple scaling argument leads us to $\eta = 2\nu_\perp - \beta$ and to the following scaling hypothesis [5]:

$$N(\tau) \sim \tau^{-\mu_\perp} f(\varepsilon^{\nu_\perp}\tau) \text{ where } \mu_\perp = 2 - \beta/\nu_\perp. \qquad (4)$$

This hypothesis first implies $N(\tau) \sim \tau^{-\mu_\perp}$ in a vicinity of the critical point ($\varepsilon \to 0$). Indeed, as shown in Fig. S7a, we observe a power-law decay of the distribution function, and the estimated exponent 1.74(2) is close to $\mu_\perp^{\mathrm{DP}} = 1.747928(7)$ derived from the above scaling relation and the theoretical value of $\beta$ and $\nu_\perp$ for (1+1)-dimensional DP. Also, integrating the hypothesis (4) from $\tau$ to $\infty$, and with a suitable normalization, we obtain the scaling hypothesis for $P(\tau)$:

$$P(\tau) = \frac{\int_\tau^\infty dt N(t)}{\int_0^\infty dt N(t)} \sim \varepsilon^{\nu_\perp(\mu_\perp - 1)} g(\varepsilon^{\nu_\perp} t) \text{ where } g(x) = \int_x^\infty dx' x'^{-\mu_\perp} f(x'). \qquad (5)$$

We rescale $P(\tau)$ according to the scaling hypothesis (5), and we observe a clear collapse onto a single, universal function as presented in Fig. S7b. Thus, we numerically confirmed that the scaling hypotheses (4) and (5) indeed hold in this case (although it may be slightly modified due to some intermittency effects [6]).

We conclude that, in a system with advection and an active boundary, one can experimentally estimate all three critical exponents $\beta$, $\nu_\parallel$ and $\nu_\perp$ by measuring an order parameter at a point sufficiently far from the boundary, spatial dependence of the order parameter, and distribution of durations of an inactive state, respectively. Note that the exponent $\mu_\perp$, which is related to $\beta$ and $\nu_\perp$ by equation (4), can be estimated from the distribution of durations of an inactive state as well.

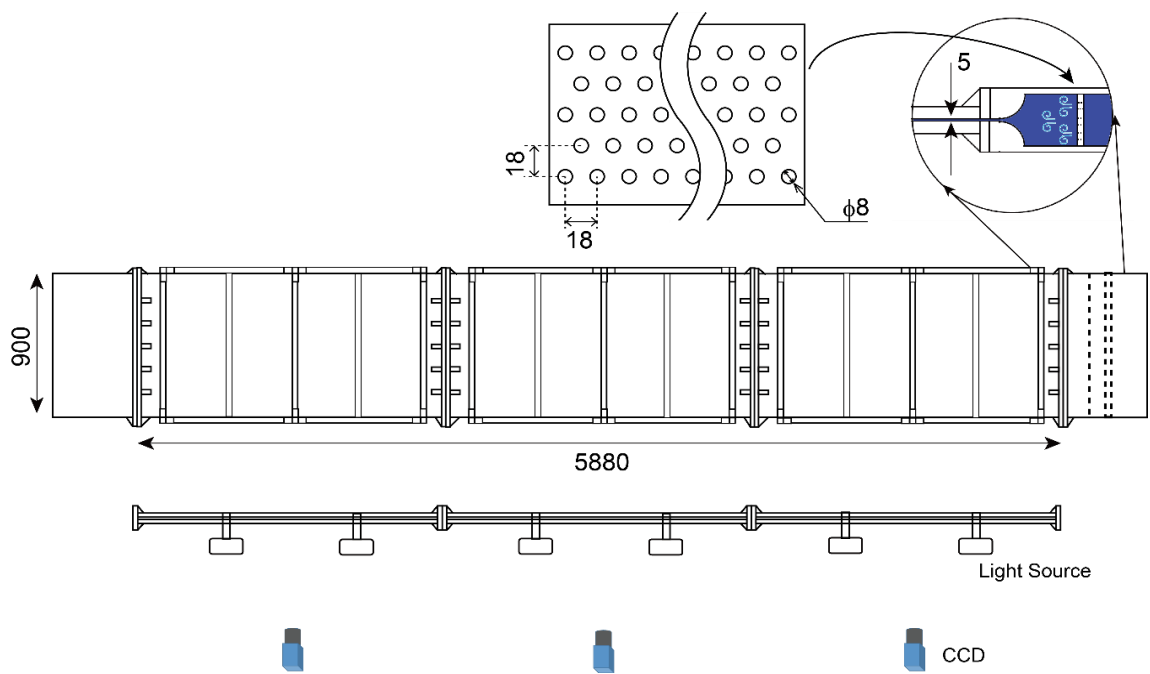

**Figure S1. Experimental apparatus.**

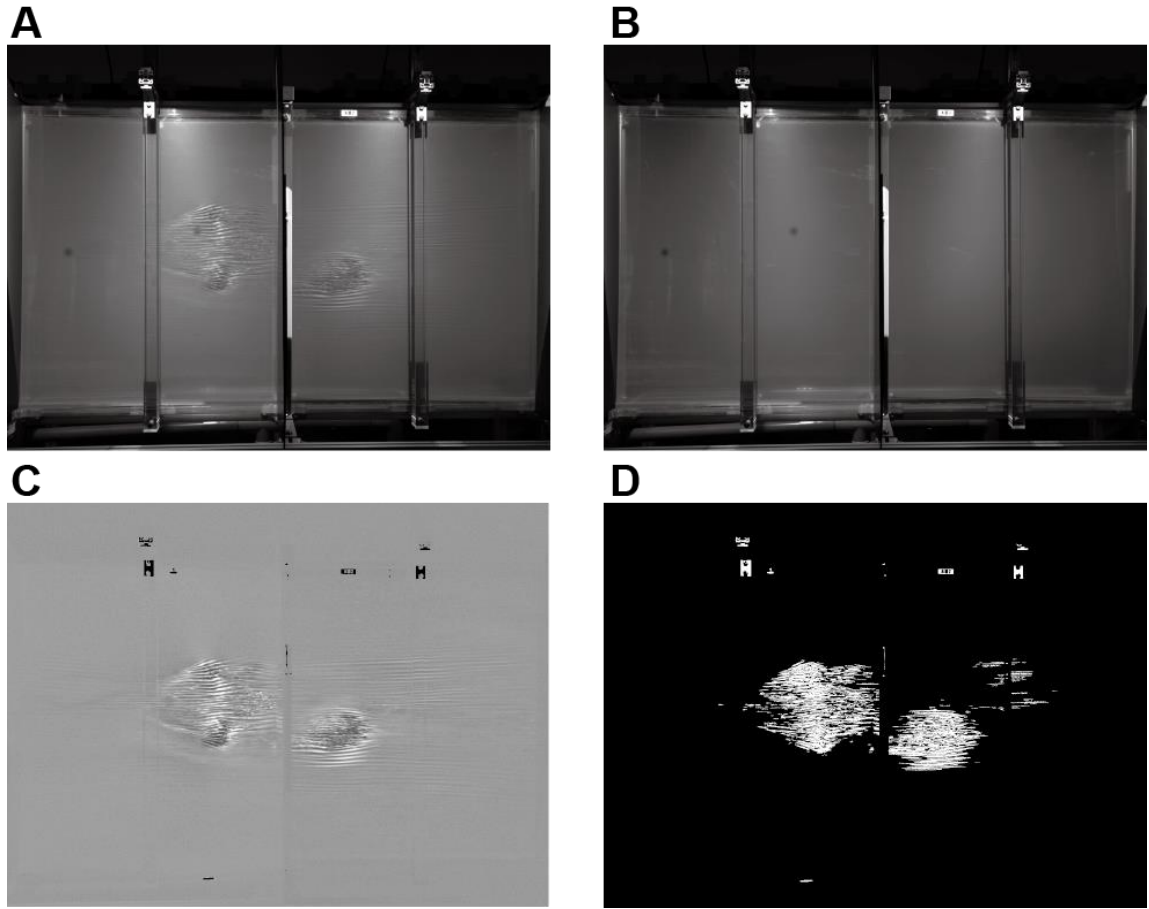

**Figure S2. Examples of image processing. a**, Original image (3 illuminating projectors are used in this sample.). **b**, Background image taken at *Re*~600 and averaged in time. **c**, Normalized image divided by the background image. **d**, Binary image showing detected turbulent regions which exceed more than 3σ(*x,z*) at each pixel.

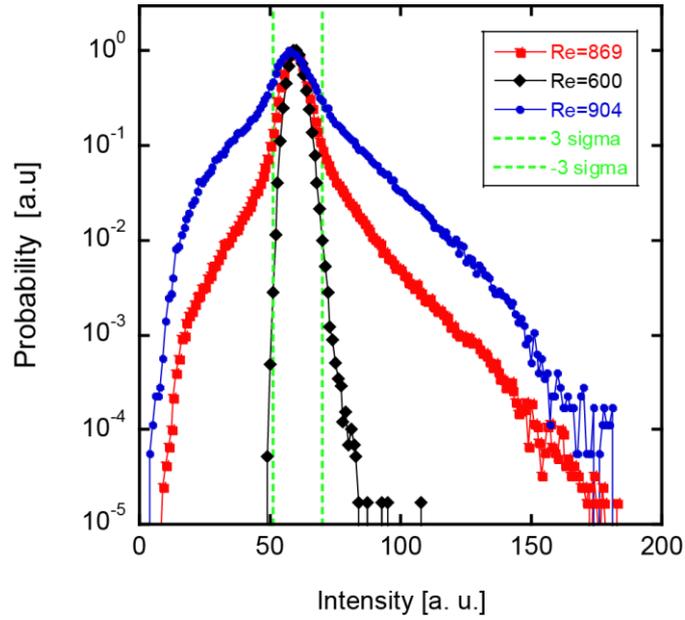

**Figure S3. Typical behavior of image intensity fluctuations.** Typical probability distribution functions (PDF) of image intensity fluctuations measured at a fixed location ($x$~3.2m) in a small region (25 mm×25mm) and accumulated for 1000 frames. All movies were normalized by a time averaged background image taken for a laminar state ($Re$~600), then histogram was accumulated for the normalized images. Black symbol represents PDF of intensity fluctuations in a laminar state ($Re$~600, ♦). PDF is close to a Gaussian distribution. As $Re$ increases to $Re$=869 (■) and $Re$=904 (●), the turbulent spots gradually increase. Accordingly, intensity fluctuations show large deviations from the Gaussian both to brighter and darker sides. Note that PDF is a superposition of a narrow Gaussian originated from laminar states and a broad distribution with large skewed wings originated from turbulent spots. Large deviations which exceed $\pm 3 \times \sigma(x,z)$ (green dashed line) from the mean laminar intensity were regarded as the turbulent state.

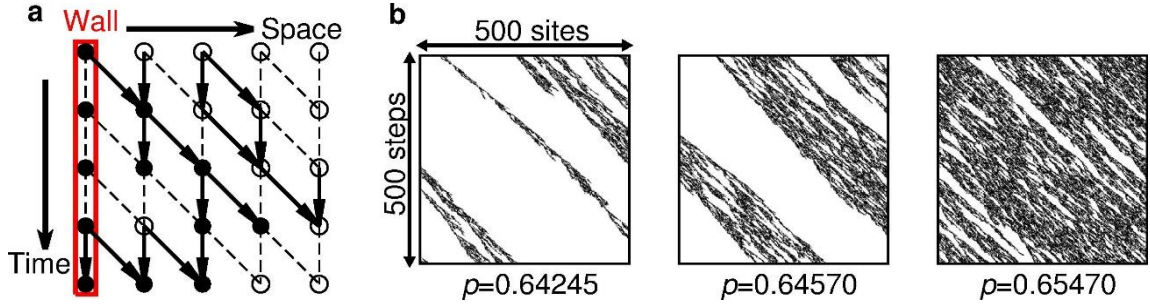

**Figure S4. Schematic picture of the (1+1)D directed bond percolation model with active wall and asymmetric connection. a**, Each site (circle) has two states, namely an active state (black) and an inactive state (white). Each bond between the sites is open (solid arrow) with probability $p$, or otherwise closed (dashed line). Sites in an "active wall" (the leftmost) are forced to be active, and sites connected with an active sites are also active. See text for a more formal definition of the model. **b**, Typical spatiotemporal dynamics observed at the 5,000—5,500 sites away from the active wall. When $p$ is smaller than the critical value ($p = 0.64245$), the clusters of active states tend to die out. When $p$ is near the critical point ($p = 0.64570$), clusters of active sites are still localized in space, but it can be sustained. When $p$ is much larger than the critical value ($p = 0.65470$), the cluster of the active sites dominate the entire space.

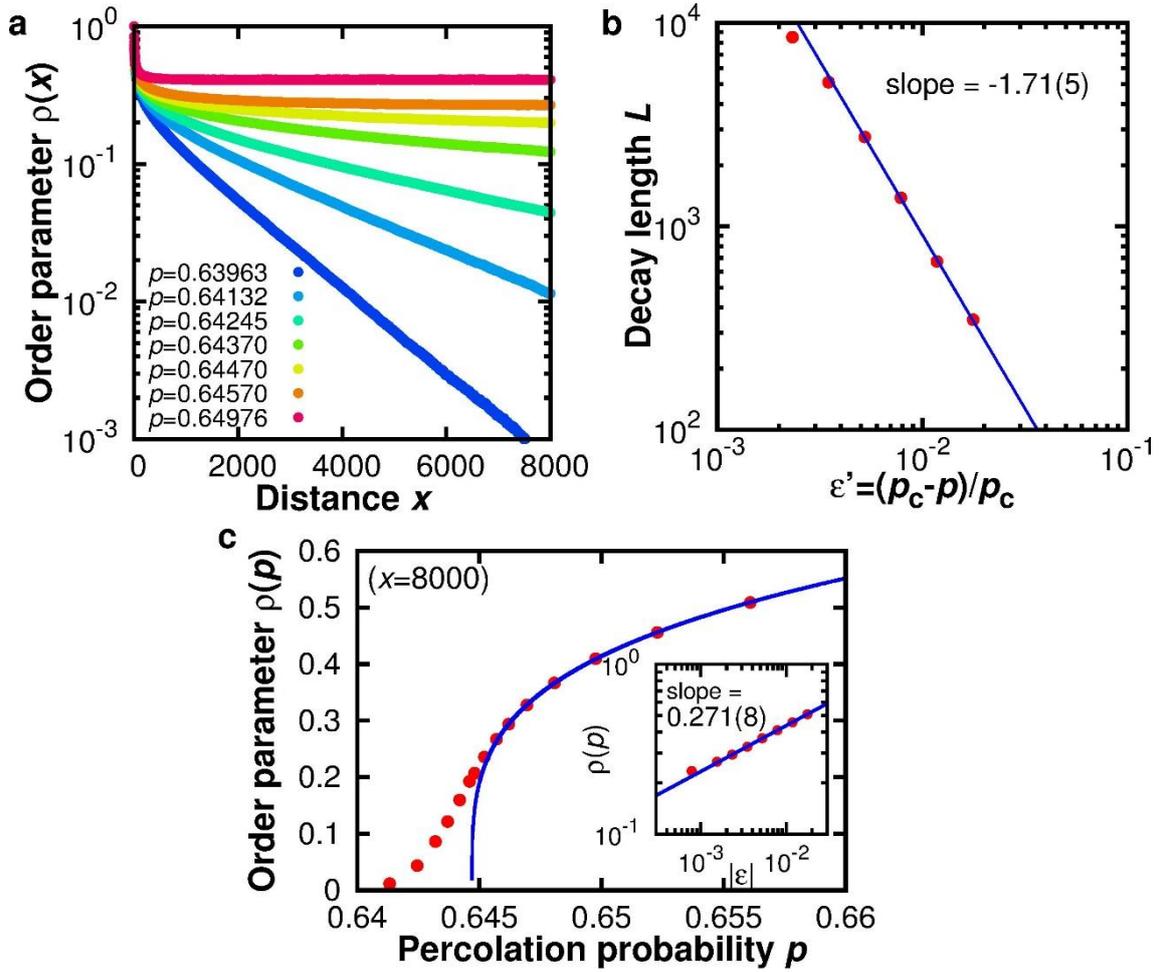

**Figure S5. Measurement of the Order parameter in (1+1)D numerical simulation. a**, Behavior of the order parameter $\rho(x)$ with respect to the distance $x$ from the active wall. When the percolation probability $p$ is significantly smaller than $p_c$, exponential decay $\rho(x) \sim \exp(-x/L)$ is observed (where $L$ is a decay length), while a convergence to a finite value is observed when $p$ is sufficiently larger than $p_c$. **b**, The decay length $L$ as a function of $\varepsilon' = (p_c - p)/p_c$. The blue solid line is a best fit by a power law. Data with $\varepsilon' > 5 \times 10^{-3}$ are used for the fitting, and the error corresponds to a 95% confidence interval in the sense of Student's *t*. **c**, Order parameter measured at 8,000[th] site from the active wall. Variation of the saturated value for $p > p_c$ is smaller than the symbol. The inset shows the same data in a logarithmic scale. A solid blue line shows the best fit (data with $\varepsilon \equiv (p - p_c)/p_c > 10^{-3}$ are used) by a power law, giving $\beta = 0.271(8)$.

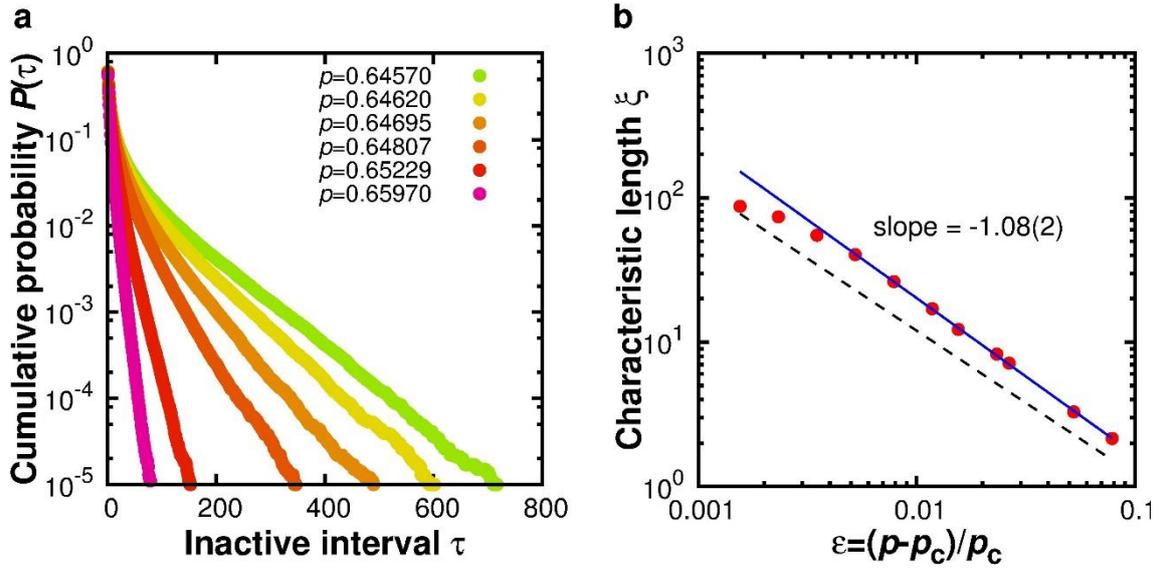

**Figure S6. Measurement of the inactive interval distribution in (1+1)D numerical simulation. a**, The probability distribution $P(\tau)$ that an inactive state is sustained up to $\tau$ steps is measured for various values of $p > p_c$, showing an exponential decay for a sufficiently large $\tau$. **b**, The characteristic length $\xi$ of the exponential decay of the distribution $P(\tau)$. The solid blue line is a best fit and the black dashed line is a guide to eye for $\xi \sim \varepsilon^{-1}$. Note that deviation from the power law in a vicinity of the critical point is observed if we measure the distribution at the point too near the active wall, where the steady state of the order parameter is not yet achieved. Similar situations occur in the experiment with turbulent liquid crystal [7] and with channel flow (see main text).

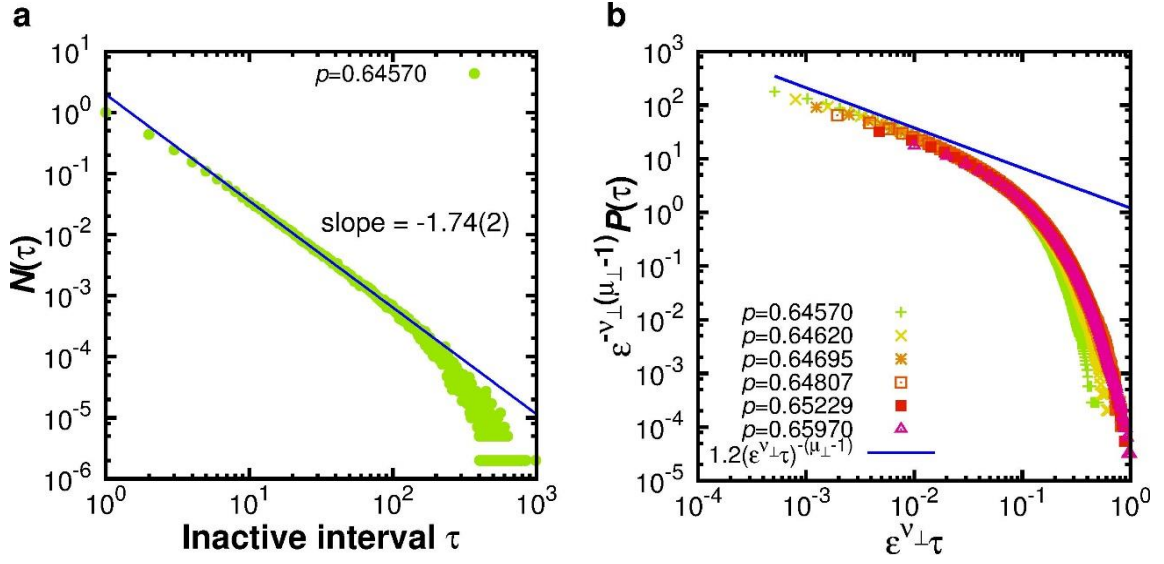

**Figure S7. Scaling property of the inactive interval distribution in (1+1)D numerical simulation. a**, The inactive interval distribution $N(\tau)$ normalized by $N(1)=1$ is shown in a logarithmic scale. Solid blue line shows the best fit, giving $\mu=1.74(2)$. **b**, The complementary cumulative distribution function $P(\tau)$ is rescaled according to the scaling hypothesis (5) (see text). Although the theoretical value of $\nu_\perp$ and $\mu_\perp$ for (1+1)D DP is used in this scaling plot, a collapse of similar quality is obtained even if we use the value estimated by the simulation. Deviation from the universal scaling function in a vicinity of the critical point is due to a finite size effect, as also observed in the experiment. Solid blue line is guide to eye for $(\varepsilon^{\nu_\perp}\tau)^{-(\mu_\perp-1)}$.